\documentclass[10pt,letterpaper]{article}
\usepackage[latin1]{inputenc}
\usepackage{opex3}
\usepackage{hyperref}
\usepackage[reqno]{amsmath}
\usepackage{amssymb}

\newcommand{\uu}[1]{\mathrm{\;#1}}
\DeclareInputMath{181}{\text{\textmu}}
\newcommand{\Nc}{\mathcal{N}}
\newcommand{\tx}{\textstyle}

\begin{document}

\title{Room temperature lasing of InAs/GaAs quantum dots in the whispering gallery modes of a silica microsphere}

\author{Sébastien Steiner\footnotemark[1], Jean Hare\footnotemark[1],
 Valérie Lefèvre-Seguin\footnotemark[1],\\ Jean-Michel Gérard\footnotemark[2]}
\address{\footnotemark[1] École normale supérieure -- Laboratoire Kastler Brossel\\
24 rue Lhomond -- 75231 Paris cedex 05 -- France }
\address{\footnotemark[2] CEA/DRFMC/SP2M -Nanophysics and Semiconductor Laboratory\\
17 rue des Martyrs -- 38054 Grenoble Cedex -- France }

\email{Jean.Hare@lkb.ens.fr} 


\begin{abstract}
We have achieved low threshold lasing of self-assembled InAs/GaAs quantum dots coupled to the evanescent wave of the high-$Q$ whispering gallery modes of a silica microsphere. In spite of high temperature and $Q$-spoiling of whispering gallery modes due to diffusion and refraction on the high index semiconductor sample, room temperature lasing is obtained with less than $100$ quantum dots. This result highlights the feasibility and interest of combining self-assembled quantum dots and microspheres in view of cavity-quantum electrodynamics  experiments.
\end{abstract}




\section{Introduction}

Semiconductor nanostructures providing efficient confinement of charge carriers have attracted an ever growing interest for several decades. Advances in semiconductor technology have allowed for the fabrication of heterostructures where bandgap discontinuities provide a binding potential in one or several dimensions. The resulting density of states is ``engineered" to provide a stronger coupling to the electromagnetic field, favorable for ever more efficient light sources. In epitaxially grown quantum dots (QDs) especially, the 3D confinement leads to the formation of ``atom like'' discrete levels, thus concentrating oscillator strength on discrete transitions. Though significant departures from a truly atom-like system have been characterized, the large oscillator strength associated with the fundamental optical transition makes QDs very promising elementary emitters for both classical and quantum light emitting devices. This transition is characterized by an optical dipole moment in excess of $10 \uu{a.u.}$ ($1 \uu{a.u.} = e a_0$) \cite{WarburtonDuerr97}, significantly greater than for atoms or ions transitions in the same spectral range, and by a moderate homogeneous broadening (down to below 1~GHz) \cite{MarzinGerard94, BayerForchel02, KammererVoisin02, BorriLangbein01}, at least at cryogenic temperatures.

QDs are therefore perfectly suited for solid-state cavity quantum electrodynamics (CQED) experiments \cite{Haroche92}, aimed at controlling spontaneous and stimulated emission properties of electron-hole pairs  by electromagnetic-mode engineering \cite{Gerard03}. In order to enhance matter--field interaction one needs microcavities with small volume modes and high quality factors. Up to now, the most common approach consists in embedding the QDs in semiconductor microcavities which can be a micropillar, a microdisk, or a photonic bandgap cavity.  The main interest of semiconductor microcavities resides in the very small mode volumes which can be achieved. Most basic CQED effects have now been observed for QDs in semiconductor cavities, including spontaneous emission rate enhancement \cite{GerardSermage98, MoreauRobert01, SantoriFattal02} or inhibition \cite{EnglundFattal05} as well as vacuum Rabi flopping for single QDs \cite{ReithmaierSek04, YoshieScherer04, PeterSenellart05}. Although quality factors as large as $300,000$ have been reported for empty microdisks \cite{SrinivasanBorselli05} or photonic crystal cavities \cite{SongNoda05}, reported values are until now significantly smaller for active semiconductor microcavities containing QDs ($Q < 30,000$).

In this context, an interesting alternative is the use of fused silica microspheres \cite{BraginskyGorodetsky89, CollotLefevre93} or microtoroids \cite{ArmaniKippenberg03} sustaining small volume whispering-gallery modes (WGM) with ultrahigh quality factors resulting from high transparency and very small residual roughness. In microspheres, the WGM are formed by successive total internal reflection along the sphere equator leading for a typical sphere of $100\uu{µm}$ in diameter to mode volumes in the range of $1000\uu{µm^3}$ and to a quality factor $Q$ as high as $10^{10}$ \cite{BraginskyGorodetsky89, CollotLefevre93}.

This unique property is particularly attractive in the context of microlaser physics and CQED. Very low-threshold ($0.2\uu{µW}$) optically pumped lasers have been demonstrated for microspheres or microtoroids doped with rare-earth ions \cite{SandoghdarTreussart96, KlitzingJahier99, LissillourFeron00,CaiPainter00-ol, YangArmani03}. Other emitters based on semiconductor heterostructures such as quantum wells \cite{FanLacey99} or nanocrystals \cite{FanPalinginis00, GotzingerMenezes06, LeThomasWoggon06} have also been coupled to silica microspheres. Such ultralow-loss cavities offer the unique opportunity to reduce the number of active emitters in a laser down to its ultimate limit \cite{ProtsenkoDomokos99}. A lasing threshold as low as 10~pA has been predicted for a microsphere coupled to a single self-assembled QD \cite{PeltonYamamoto99}. Although such a laser would only work at low temperature \cite{Gerard03}, it is obviously a very attractive model system to study experimentally the very peculiar physics of single ``atom'' lasers in a solid-state system.

In this letter, we report for the first time the laser operation at room temperature of a silica microsphere coupled to few InAs/GaAs self-organized QDs used as the amplifier medium. We had to overcome two challenging problems: the poor radiative efficiency of the QDs at room temperature resulting from thermal promotion of carriers to the continuum states of the wetting layer, and the strong broadening of WGMs induced by the high refractive index sample. Despite these difficulties, we nevertheless achieved cw lasing with thresholds as low as $200\uu{µW}$ in absorbed pump power. Working with microstructured samples was a key factor for this result.

\section{Principle of experiment}

\subsection{Main features of Microsphere WGMs}
The WGMs of dielectric microspheres are well known and detailed structural and spectroscopic properties can be found in \cite{OpticalProcesses96} and reference therein. Here we describe briefly only the main features of interest for the experiment. The high-$Q$ WGMs of a silica microsphere of radius $a\sim 10-100\uu{µm}$ and refractive index $N\approx 1.45$ are described, beside their TE or TM polarization, by 3 integer numbers $n$, $\ell$, $m$ characterizing the electric field distribution $E \propto f_{n\,\ell}(r) Y_\ell^m(\theta,\varphi)$, where $(r,\theta,\varphi)$ are the spherical coordinates. While $n$ is the number of antinodes of the radial function $f_{n\,\ell}(r)$, $\ell$ and $m$ are the usual angular ``quantum" numbers, associated to $\ell-|m|+1$ anti-nodes in the polar direction, as experimentally observed in Ref.~\cite{KnightDubreuil95}. The modes with $n = 1$ and $|m| = \ell$ are confined at the wavelength scale around the sphere equator, with an evanescent wave in the surrounding air medium roughly decaying as $\exp\big(-\kappa (r-a)\big)$, where $\kappa\approx (N_\text{eff}^2-1)^{1/2}\, 2\pi/\lambda$, with  $N_\text{eff}=\ell\lambda/2\pi a$ the WGM's effective index, close to the  refractive index of silica $N$.

The WGMs therefore verify the relation $2\pi a = \ell\lambda/N_\text{eff}$, and this results in a quasi periodic spectrum with an effective free spectral range (FSR) $\Delta\nu_\text{\,FSR}\approx c/2\pi N_\text{eff} a$.
Due to a small ellipticity, the WGMs differing only in $|m|$ exhibit different resonance frequencies. In a given FSR, frequency scanning combined with appropriate beam settings allows us to excite modes of various $n$ or $\ell-|m|$, which can be identified by spectroscopic methods.

\subsection{QD Sample}
We used a sample containing InAs/GaAs QDs grown by molecular beam epitaxy on a (001) GaAs substrate. The epitaxial structure consisted in a 500~nm thick GaAs buffer layer, 30~nm $\mathrm{Al_{0.15}Ga_{0.85}As}$, 50~nm GaAs, a layer of InAs QDs, 20~nm GaAs, 10~nm $\mathrm{Al_{0.15}Ga_{0.85}As}$, and on top a 2.1~nm GaAs cap layer. The $\mathrm{Al_{0.15}Ga_{0.85}As}$ layers act as barriers for electrons and holes and prevent their diffusion toward the surface, which is known as a major non-radiative recombination center in GaAs based materials. The capping layer grown on top on the QDs was kept as thin as possible in order to maximize the evanescent coupling to the WGM. The sample was furthermore processed by electron-beam lithography and reactive ion etching so as to define an array of  $4 × 4\uu{µm}$ micromesas about 200~nm in height, $40\uu{µm}$ apart from one another.

Microphotoluminescence experiments have been performed in order to characterize this QD array under pumping by a focused laser diode at $\lambda_\text{pump} = 778\uu{nm}$ (1.59~eV) with a power of typically $1\uu{mW}$ within a $5\uu{µm}$ spot diameter. At room temperature the QD array exhibits a broad emission line centered at $\lambda_\text{PL} = 1075\uu{nm}$ (1.15~eV). Its linewidth at half maximum of 100~nm (ie $25\uu{THz}$ or 110~meV) is mainly due to the inhomogeneous broadening resulting from size dispersion, while homogenous broadening due to phonon scattering is known to be about 2.5~THz (10~meV) at 300~K \cite{BorriLangbein99}. At 8~K on the other hand, and under weak excitation conditions (in the 1 to $10 \uu{µW}$ range), the QD emission spectrum was observed to consist in a series of sharp lines, being each related to the emission of one specific QD. Using this feature, we determine the number $\Nc$ of QDs per mesa ($\Nc\simeq 600$) as well as their spectral distribution for the mesas of interest.

\subsection{Experimental setup}

\begin{figure}[htbp]
 \centering
 \includegraphics[width= 100mm]{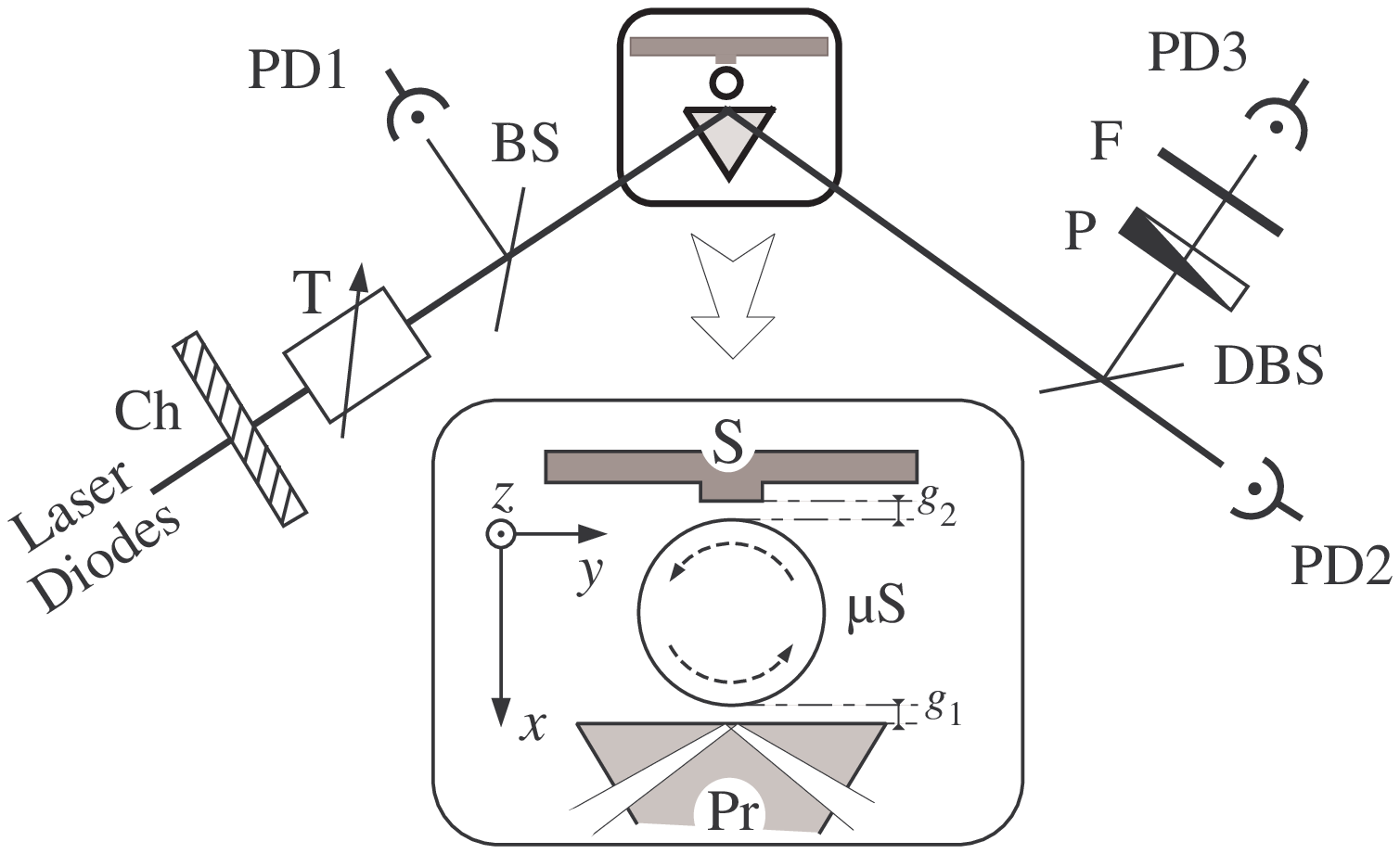}
 \caption{Scheme of our experimental setup (not to scale). \textmu S: microsphere; Pr: SF11 glass coupling prism; S: sample; Ch : chopper; T : attenuator; BS : beam-splitter; DBS : dichroic beam-splitter; P:~polarization analyzer; F:~Wratten 87C or Schott RG1000 filter. PD1/2: silicon photodiodes for pump monitoring, PD3: InGaAs photodiode for QDs emission monitoring. Bottom inset: WGM coupling geometry, showing the excitation prism Pr, the spherical microcavity \textmu S and the sample S. }
 \label{f:setup}
\end{figure}

The experimental setup is sketched in Fig.~\ref{f:setup}. It has been described in previous work (see e.g. Ref.~\cite{SandoghdarTreussart96}). The main advantage of prism-coupling over alternative techniques is the ability to optimize input coupling of the pump while keeping maximal output coupling of the photoluminescence (PL), whatever the wavelength and the involved modes.

In our experiment, the microsphere is placed between the SF11 glass prism and the GaAs sample. This configuration has several key advantages: the pumping laser at $\lambda_\text{pump}\sim 780\uu{nm}$ is coupled resonantly to a WGM  ensuring that only the QDs actually coupled to it can be excited, and that the PL extracted through the prism has necessarily been emitted inside WGMs.
The drawback is that the microsphere is almost stuck between two millimeter-sized objects which makes more difficult the overall alignment. The parallelism of the sphere's stem with respect to the prism and the sample is of course crucial in order to prevent any unwanted contact. One also has to control independently the two coupling gaps: $g_1$ between the prism and the sphere, and $g_2$ between the sphere and the micromesa under study, as shown in the inset of Fig.~\ref{f:setup}.  Our setup is designed to provide control over the overall 18 degrees of freedom with the necessary precision, using micrometer screws and piezoelectric transducers (PZT).

We used three different laser diodes for this experiment. Beside the pump laser (see §.3), two DBR laser diodes operating at $\lambda_1 = 772\uu{nm}$ and $\lambda_2 = 1,060\uu{nm}$ served to characterize the prism-sphere and sphere-sample coupling parameters at both the pumping and the emission wavelengths, and to measure the cold-cavity linewidth.


\subsection{Sphere--Sample coupling optimization}

\begin{figure}[htbp]
 \centering
 \includegraphics[width= 100mm]{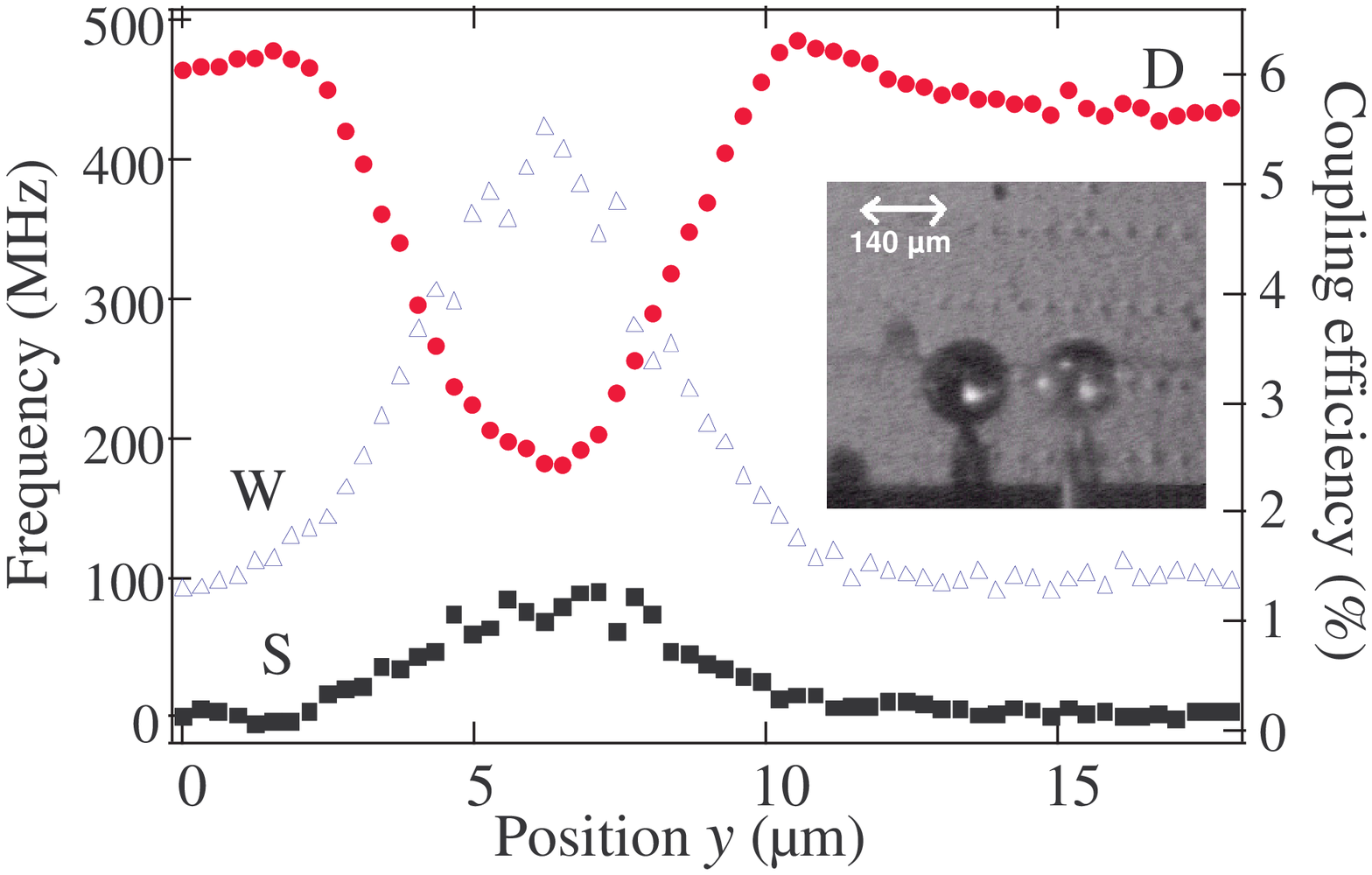}
 \caption{Evolution of the width ($\scriptstyle\blacksquare$), shift ($\color{blue}\scriptstyle\bigtriangleup$) and coupling ($\color{red}\bullet$) when the sample is scanned along $y$ for arbitrarily fixed $g_1$ and $g_2$ (see Fig.~\protect\ref{f:setup} for axes and gaps definitions).
Inset : Photograph of the sphere (diameter $140\uu{µm}$) in front of the microstructured sample. The sphere is on the right and the image to the left is its reflection in the sample. The small dots are $4\uu{µm} × 4\uu{µm}$ mesas, 200~nm in height, revealed by grazing illumination.}
 \label{f:optim}
\end{figure}

Positioning a given mesa in the field of a strongly confined WGM is absolutely critical. A rough alignment of the setup with a given mesa is first achieved using a stereomicroscope, as shown in the inset of Fig.~\ref{f:optim}, but visual inspection does not work for further optimization. In order to PZT control the positioning of the selected mesa with the desired precision, we monitor the influence of the mesa directly on the WGM resonance signal (amplitude, shift and broadening).

We indeed observe significant sample-induced broadening of the WGM resonances, but no more than the usual broadening due to the prism.  With a simple model describing the perturbation induced by a plane dielectric sitting in the WGM evanescent field, we have shown (see Ref.~\cite{Treussart_phd, Steiner}) that this modification is characterized by the Fresnel reflection coefficient for the evanescent wave which for TE modes writes:
\begin{equation}\label{eq:R}
   r\ \approx\ \tx\frac{ i\sqrt{\tx N_{\text{eff}}^2-1}-\sqrt{\tx N_D^2-N_{\text{eff}}^2}}%
                {i\sqrt{\tx N_{\text{eff}}^2-1}+\sqrt{\tx N_D^2-N_{\text{eff}}^2}} \ ,
\end{equation}
in which $N_\text{eff}$ is the WGM effective index, and $N_D$ the (eventually complex) index of the dielectric. This reflection coefficient accounts for a frequency shift proportional to the real part of $r$, and a line broadening proportional to its imaginary part. With $N_\text{eff}\approx N \approx 1.45$ and $N_D\approx 1.76$ for SF11 or $N_D\approx 3.36$ for GaAs (at 1,060~nm), one gets $r\approx 0.05 + i\, 0.999$ for the prism, and $r\approx0.77 + i\, 0.63$ for GaAs. This means that the index of SF11 gives an almost vanishing shift and a large broadening, as already confirmed in earlier work \cite{DubreuilKnight95}. On the other hand, the high index of GaAs results in a large shift and a rather moderate broadening \cite{note1}.  We have experimentally confirmed these predictions using a flat (un-etched) pure GaAs sample. At zero-gap, the WGMs efficiently coupled to this flat sample exhibit linewidths in excess of 10~GHz. This large broadening was a strong motivation to work with micromesas instead of a flat QD sample.

As an example of our positioning method, Fig.~\ref{f:optim} shows the data obtained when scanning the position of a mesa in a direction $y$ approximately parallel to the equator, with the gaps $g_1$ and $g_2$ (along the $x$-axis) and the height ($z$ coordinate) kept constants (see Fig.~\ref{f:setup} for axes definition). Because the length scales of $g_{1,2}$, $y$ and $z$ are much smaller than the sphere radius $a$, a parabolic expansion $r-a\approx g_2+(y^2+z^2)/2a$, transforms the exponential behavior of the WGM evanescent wave into a Gaussian dependence $\exp(-\kappa y^2/a)$ for its intensity versus $y$, where $\kappa\simeq (N_\text{eff}^2-1)^{1/2}\, 2\pi/\lambda$. Because the height of the mesa ($\sim 200\uu{nm}$) is larger than $\kappa^{-1}$, and its width larger than the transverse field extension $(a/\kappa)^{1/2}$, the mesa can be seen in a first approximation as a small area sample. The behavior of the width or the shift of the line thus appears as a convolution of the square mesa profile and the Gaussian, with an expected full width at half maximum of about 5~µm, in good agreement with our data \cite{note2}. Due to a slight tilt (about 0.5\textdegree) between the sphere axis and the vertical axis $z$, the best coupled mode was a mode with $m =\ell-7$. The records obtained for this mode when $z$ is scanned (not shown here) actually reveal two antinodes of the polar intensity distribution.
In these scans, the observed shift/broadening ratio of 28\% is much smaller than the expected value of 78\%, as given by Eq.~\ref{eq:R}. This reveals an extra loss mechanism attributed to strong light scattering at the edges of the mesa, which will be taken into account below to estimate the pumping efficiency.

\section{Laser operation}

\begin{figure}[htbp]
 \centering
 \includegraphics[width= 90mm]{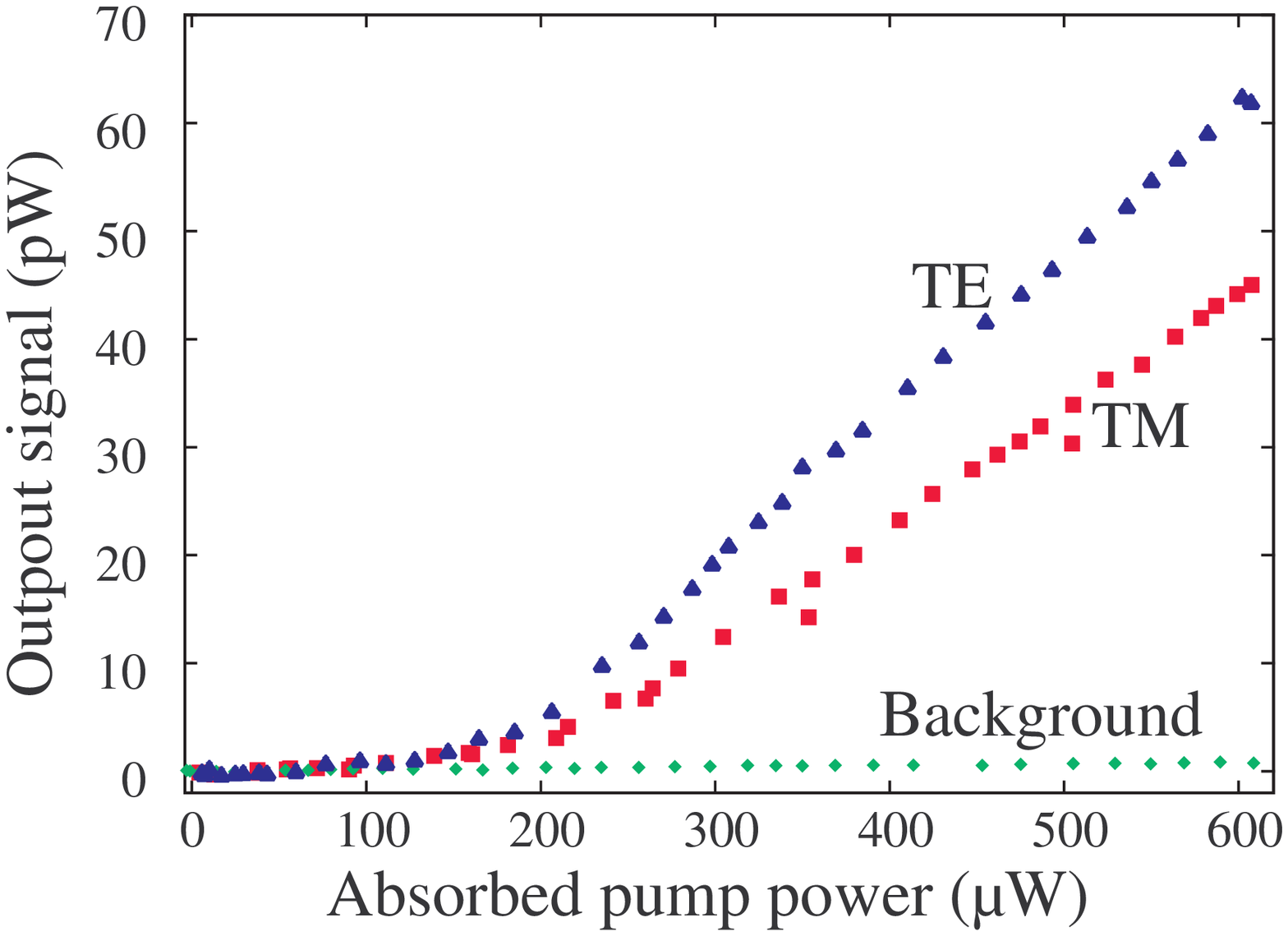}
 \caption{Emission vs. absorption laser characteristics for both WGM polarizations. The background curve shows the residual fraction of pump laser fluorescence at wavelength higher than 900 nm.}
 \label{f:laser}
\end{figure}

The pump laser used to achieve QD lasing was a widely tunable Littman-Metcalf extended-cavity laser diode (Newport/EOSI 2010) with a central wavelength of about 780~nm. Special care had to be taken to filter out a small fluorescence from this laser-diode in the 900--1,100~nm range. This parasitic signal which otherwise hides the QD's emission is reduced by about two orders of magnitude thanks to a dispersion prism inserted at the exit of the pump laser. The background plotted in diamonds in Fig.~\ref{f:laser} corresponds to its residual level, as measured when the sample is wide apart from the sphere.

The QDs emission, outcoupled from the WGMs via the prism Pr, was monitored using an InGaAs photodiode (Hamamatsu G3476-10) providing sensitivity at the pW level (PD3 in Fig.~\ref{f:setup}).

To extract the PL signal from the noise, the pump beam was chopped at a frequency of about 200~Hz, and a lock-in amplifier was used. We also inserted in front of PD3 three sheets of gelatine IR filter (Kodak Wratten 87C) to further eliminate the residual pump light.

The pump laser was tuned to a strongly confined WGM chosen to maximize the coupling efficiency to the QDs. The observed shift/broadening ratio indicates that about one third of the pump power contributes to the photoexcitation of the sample (above the bandgap of GaAs).

The sphere--prism  and sphere--sample gaps (resp. $g_1$ and $g_2$) also have to be optimized.
Because the QDs active medium--field coupling and the resonance broadening due to the mesa exhibit the same exponential behavior versus $g_2$, there is  \textit{a priori} no preferred $g_2$ value for laser operation. However, a simple model taking into account the losses due to the prism shows that the lowest threshold should be obtained when $g_2 = 0$. We therefore worked with the sample in contact with the sphere, a setting which furthermore provides much better mechanical stability. This was made possible only because the Q-spoiling due to the microstructured sample is strongly reduced by comparison with a flat QD sample. With the cavity thus loaded by the sample, the sphere--prism gap $g_1$ was set close to critical coupling. In this situation, the measured cold-cavity width of the best coupled WGMs at $1,060\uu{nm}$ was $\Gamma/2\pi = 1\uu{GHz}$, corresponding to $Q=0.3×10^6$, with equal weight for the sample and the prism.

With these settings it was possible to observe the onset of lasing evidenced by a well pronounced threshold in the pump/emission characteristics. Figure \ref{f:laser} shows such a laser characteristic curve for both TE and TM polarization, with a threshold of $200\uu{µW}$, the lowest observed. In this figure, the ``output signal'' is the raw detected signal, not corrected for collection and detection efficiency. The ``absorbed pump power'' is the power lost from the incident beam into the sphere (including scattering) as monitored by photodiodes PD1/PD2 in Fig.~\ref{f:setup}. These curves were obtained for a $140\uu{µm}$ sphere, pumped into a $n = 1$, $m = \ell-6$, TE polarized WGM.

Due to the small output power and to the lack of a low-noise cooled spectroscopic camera, it was not possible to record a spectrum of this microlaser. However when inserting additional sheets of Wratten 87C filter or a Schott RG1000 filter, we observed a very weak attenuation of the output signal. This demonstrates that the wavelength of the emitted light is 1,000~nm or higher. In particular, the RG1000 filter spectral response gives an attenuation 10 times stronger for the GaAs or InAs wetting layer emission than for the QDs emission in the $1075\pm50\uu{nm}$ range.

Since the coupling of the mesa to the sphere was maximal for a WGM with $n = 1$ and $m =\ell-6$, the lasing modes also have $n=1$ and $\ell-|m|\approx 5$ providing the best overlap with the mesa. The $\ell$ value is not limited by the same overlap constraints, and can a priori cover the gain spectral range, hence leading to $\ell$ values lying between 550 and 600. Multimode operation is therefore very likely, in spite of mode competition.

\section{Discussion}

\subsection{Pumping efficiency}

We first discuss the pumping rate achieved in our system.
The pumping efficiency is strongly limited by several phenomena. We first take into account the pumping efficiency reduction to about  30\% due to the scattering effect revealed by the shift/broadening ratio. Besides, the power actually injected in the sample is absorbed on a characteristic length of about $1\uu{µm}$, whereas the useful zone between the AlGaAs barriers is only 70~nm thick : this introduces an additional factor of 7\%, resulting in an overall efficiency of about 2\%. An absorbed pumped power of $200\uu{µW}$ therefore  corresponds to about $15×10^{12}$ electron-hole pairs per second.  When trapped in a QD, these electron-hole pairs have a total lifetime of about 100~ps at room temperature. This shows that the absorbed pump power is large enough to feed the $\Nc\approx 600$ QDs in a mesa.

\subsection{Threshold condition}

The experimental parameters of our system can be used to discuss the lasing threshold condition for a given WGM. The average photon number at threshold $n^\text{(th)}$ verifies:
\begin{equation}\label{e:threshold}
    n^\text{(th)} = \Nc_c^\text{(th)} \frac{W}{\Gamma} \approx 1\quad \Rightarrow \quad \Nc_c^\text{(th)}\ \frac{\Omega_R^2}{\gamma_\text{hom}\Gamma} \approx 1 ,
\end{equation}
with $\Nc_c$ the number of QDs coupled to the mode, $\Omega_R$  the vacuum Rabi frequency, $W$ the spontaneous emission rate in the mode, $\Gamma/2\pi$ the WGM's linewidth and $\gamma_\text{hom}/2\pi$ the QD homogenous linewidth. Here, as shown in \cite{ProtsenkoDomokos99}, the second equality results from $W = \Omega_R^2 /(\gamma_\text{hom}+\Gamma)\approx  \Omega_R^2 /\gamma_\text{hom}$,  because here $\gamma_\text{hom}\gg\Gamma$ (good cavity regime). In the experiment discussed here the WGM mode volume is  about $5,000\uu{µm^3}$ in the lasing wavelength range, leading to a maximal Rabi frequency $\Omega_R/2\pi\approx 2~\uu{GHz}$ at the point where the WGM's field reaches its maximum value $E_\text{max}$.

Using the above mentioned parameters, we are left with the frequency hierarchy:
 \begin{equation}
 \frac{\gamma_\text{inh}}{2\pi} \approx 25,000\uu{GHz} \gg
 \frac{\gamma_\text{hom}}{2\pi} \approx 2,500\uu{GHz} \gg
\frac{\Delta\omega_\text{\ FSR}}{2\pi} \approx 500\uu{GHz} \gg
\frac{\Gamma}{2\pi} \approx 1\uu{GHz}
 \end{equation}
for the QD's inhomogeneous width, the QD's homogenous width, the WGM's free spectral range and the WGM linewidth, respectively.

As stated earlier, the recombination emission of the $\Nc$ QDs is spread over a frequency span $\gamma_\text{inh}/2\pi$, which encompasses about 50 FSR. Inside a $\gamma_\text{hom}/2\pi$ frequency bandwidth, mode competition is expected to favor one lasing mode. Such a mode interacts resonantly with $\Nc_c\approx\Nc\gamma_\text{hom}/\gamma_\text{inh} \approx 60$ QDs.

Considering the critical parameter $\Omega_R^2/\gamma_\text{hom}\Gamma$ which enters Eq.~\ref{e:threshold}, it is of the order of $1.6×10^{-2}×(E_\text{QD}/E_\text{max})^2$.  The threshold condition can be fulfilled if the ratio $(E_\text{QD}/E_\text{max})^2$ is close enough to 1. This condition suggests that the WGM field in the coupling zone is deconfined from the sphere to the high index sample. We believe that this reconstruction mechanism explains our observations when studying different mesas: although they show the same PL characteristics, some lase while others do not, everything else being equal. Indeed the mode reconstruction, which depends on the precise size and structure of the mesa, is expected to vary from one mesa to the other.

To summarize, the orders of magnitude discussed above are consistent with an absorbed pump power of $200\uu{µW}$ at threshold sustaining about 10 modes with about 60 active QDs per mode.

\section{Conclusion}

We have successfully combined the high quality factor of the WGMs in a microsphere and the large oscillator strength of III-V semiconductor QDs to achieve room-temperature lasing with less than 100 QDs per mode. Working with QDs in micromesas instead of a flat sample was a decisive improvement. The results reported here clearly demonstrates the potential of high-$Q$ WGM microcavities coupled to self assembled quantum dots. Further enhancement of QD--WGM field coupling efficiency, as measured by the spontaneous emission rate $W$, relies on smaller homogenous linewidths $\gamma_\text{hom}$ and minimized $Q$-spoiling. Decreasing the homogenous linewidth by a factor of about 1000 or more could be achieved by a cryogenic version of the experiment \cite{TreussartIlchenko98}, in order to get $\gamma_\text{hom}\approx\Omega_R$. The preservation of microsphere ultrahigh-$Q$ requires smaller mesa sizes and no bulk sample substrate. In this perspective, the ``van~der~Waals bonding'' on the sphere of a microsample of $1\uu{µm}$ in diameter and $100\uu{nm}$ in height has been undertaken. This procedure can furthermore make much easier the cryogenic experiment. When achieved, this enhanced version of the experiment reported here will allow to reach the thresholdless laser regime ($\Nc_c^\text{(th)}\approx 1)$, as characterized theoretically in \cite{ProtsenkoDomokos99}, and to study the ``one-atom'' laser with a single QD coupled to a microsphere.

\section*{Acknowledgements}
The authors warmly thank Jean-Michel Raimond and Serge Haroche for fruitful and encouraging discussions.
JMG gratefully thanks B. Gayral and Y. Nowicki for the characterisation of the QD sample by microphotoluminescence.
Laboratoire Kastler Brossel is a laboratory of École normale supérieure and Université P.\&M.~Curie, associated to C.N.R.S. (UMR 8552). This work was partly supported by CNRS-Japan Science and Technology contract (Stanford/Ens ICORP ``Quantum Entanglement" Project) and by European Community (QUBITS Network).

\end{document}